\documentclass[9pt,twocolumn,twoside]{pnas-new}
\usepackage{CJKutf8}

\templatetype{pnasbriefreport} 

\def\iso#1#2{\mbox{${}^{#2}{\rm #1}$}}
\def\o1#1{\iso{O}{1#1}}
\def\fe6#1{\iso{Fe}{6#1}}
\def\sm14#1{\iso{Sm}{14#1}}
\def\u23#1{\iso{U}{23#1}}
\def\pu24#1{\iso{Pu}{24#1}}

\title{Do we Owe our Existence to Gravitational Waves?}

\author[a,1]{John Ellis}
\author[b]{Brian D. Fields}
\author[c]{Rebecca Surman}

\affil[a]{Department of Physics, Kings College London, Strand, London WC2R 2LS, UK;
\\
Theoretical Physics Department, CERN, CH-1211 Geneva 23, Switzerland}
\affil[b]{Department of Astronomy, Illinois Center for Advanced Studies of the Universe, and  Department of Physics, University of Illinois, 1002 W. Green St., Urbana IL 61801, USA}
\affil[c]{Department of Physics and Astronomy, University of Notre Dame, Notre Dame, IN 46556, USA}

\leadauthor{Ellis} 

\authorcontributions{Author contributions: .}
\authordeclaration{The authors are unaware of any competing interests.}
\correspondingauthor{\textsuperscript{1}To whom correspondence should be addressed. E-mail: john.ellis@cern.ch}


\begin{abstract}
Two heavy elements essential to human biology are thought to have been produced by the astrophysical $r$-process, which occurs in neutron-rich environments: iodine is a constituent of thyroid hormones that affect many physiological processes including growth and development, body temperature and heart rate, and bromine is essential for tissue development and architecture. Collisions of neutron stars (kilonovae) have been identified as sources of $r$-process elements including tellurium, which is adjacent to iodine in the periodic table, and lanthanides. Neutron-star collisions arise from energy loss due to gravitational-wave emission from binary systems, leading us to suggest that gravitational waves have played a key role in enabling human life by producing iodine and bromine. We propose probing this proposal by searching in lunar material for live $^{129}$I deposited by a recent nearby kilonova explosion.
\end{abstract}

\dates{This manuscript was compiled on \today}
\doi{\url{www.pnas.org/cgi/doi/10.1073/pnas.XXXXXXXXXX}}

\begin{document}

\maketitle
\thispagestyle{firststyle}
\ifthenelse{\boolean{shortarticle}}{\ifthenelse{\boolean{singlecolumn}}{\abscontentformatted}{\abscontent}}{}

\section*{Introduction}

Ever since observations of `guest stars' were first recorded~\cite{xu2000east}, there has been widespread interest in their possible significance as harbingers of life and death. Many of the most important elements in the human body, such as carbon and oxygen, originated {in large part} from supernova explosions~\cite{Hoyle1946}.  
Supernovae are thus essential for life, yet explosions too nearby may have caused
one or more mass extinctions of life on Earth~\cite{ruderman1974possible,Ellis1993}. 
This motivated the search for nuclear abundance anomalies in geological strata from the time when the dinosaurs met their demise, which found the first evidence that their extinction was in fact due to an asteroid impact~\cite{alvarez1980extraterrestrial}. Detailed modeling~\cite{Thomas_2023} indicates that a supernova explosion within $\sim 20$~pc of Earth, which is expected to occur every few hundred million years, would cause a mass extinction. 

In recent years there has been increasing interest in neutron-star collisions (kilonovae)~\cite{metzger_kilonovae_2020}. These events are estimated to have a similar `kill radius' to supernovae~\cite{perkins2023kilonova} unless the Earth is unlucky enough to be within the narrow line of sight of a Gamma-Ray Burst emitted by the kilonova, but are much rarer than supernovae and hence less dangerous. In contrast to these existential dangers, we focus in this article on the benefits to life of kilonova explosions, arguing that they produce elements essential to human life, and that they are made possible by gravitational waves like those discovered recently~\cite{LIGOScientific:2017vwq}. In order to make this argument, we first review relevant aspects of human biochemistry.

\section*{The Roles of Heavy Elements in Human Biology}

The human body is largely composed of the elements hydrogen, carbon and oxygen, with smaller amounts of other elements. The total number of elements adjudged essential for life exceeds 20. Most of these have atomic numbers $Z < 35$ and are produced by supernovae~\cite{Hoyle1946}. However, two heavier essential elements stand out that are produced mainly by the astrophysical $r$-process that we discuss below: bromine with $Z = 35$ and iodine with $Z = 53$. 

Iodine is the better known of this pair: the thyroid hormones triiodothyronine and thyroxine play key roles in many physiological processes in the human body, including growth and development, metabolism, and controlling body temperature and heart rate~\cite{Nussey,NRC}. The role of bromine is less commonly known, but it is is an essential trace element for the assembly of collagen IV scaffolds in tissue development and architecture~\cite{mccall2014bromine}.

We mention in passing some other heavy elements. Molybdenum ($Z = 43$)
is a key component of enzymes that promote oxygen atom transfer reactions~\cite{kisker1997molybdenum,NRC} 
that are present in the mitochondria of all eukaryotes and important for the generation of ATP, a key cellular energy source.
It has also been suggested that cadmium ($Z = 48$)
Finally, we note that 
the 
gadolinium and uranium ($Z = 64, 92$) play roles in the biochemistry of some bacteria~\cite{daumann2019essential},
though there is no evidence for any biological action of these elements in mammalian biology.

It is also possible that thorium and uranium have been indirectly important for human life. Plate tectonics has played an important role in the evolution of life on Earth, and possibly could be similarly important on Earth-like exoplanets \cite{Korenaga2012,Frank2014,Noack2014}.  Tectonic activity is possible because the Earth's interior remains hot enough to enable plate motion. 
The decays of radioisotopes \iso{U}{235,238} and \iso{Th}{232}  in the Earth's interior generate $\gtrsim 50\%$ of geothermal heat today, a fact confirmed by measurements of geoneutrinos also produced in these decays~\cite{Borexino2020}.

In the following we focus on iodine and bromine and their astrophysical origins.

\section*{Heavy Element Production via the $r$-Process}

The rapid neutron-capture process, known as the $r$-process, plays an important role in the production of atomic nuclei heavier than iron. Seed nuclei capture a succession of neutrons in time to avoid radioactive decay before another neutron is captured. The reaction pathway therefore lies far from stability, possibly out the neutron drip line, beyond which the nuclear force can no longer retain neutrons. The $r$-process yields characteristic abundance peaks at $A\sim 80$, 130, and 195, associated with the closed nuclear shells at the magic neutron numbers $N=50$, 82, and 126. Whether all three abundance peaks owe their origins to the same astrophysical site is an open question, e.g. \cite{Roederer2023}. However, species within the same abundance peak, such as the elements tellurium ($Z=52)$ and iodine ($Z=53)$, are likely co-produced, with a ratio set by the properties of the unstable nuclear species along the relevant $r$-process pathway \cite{Mumpower+2016}.

The $r$-process is not the only source of heavy elements in the solar system.
The main other source is from slow neutron capture via the so-called {\em s}-process.
In that case the nuclei remain close to stability, so the nuclear physics is very well understood, so that 
there are relatively tight predictions for solar system {\em s}-process abundances \cite{Kappeler+2011}.
The astrophysical sites are dominated by the late phases of stars with masses above that of the Sun but less than that of supernova progenitors. These stars generate neutrons and forge heavy elements in the
asymptotic giant branch (AGB) phase that precedes the ejection of the newly-synthesized elements as a planetary nebula. 

\begin{figure}
    \centering
    \includegraphics[height=0.4\textwidth]{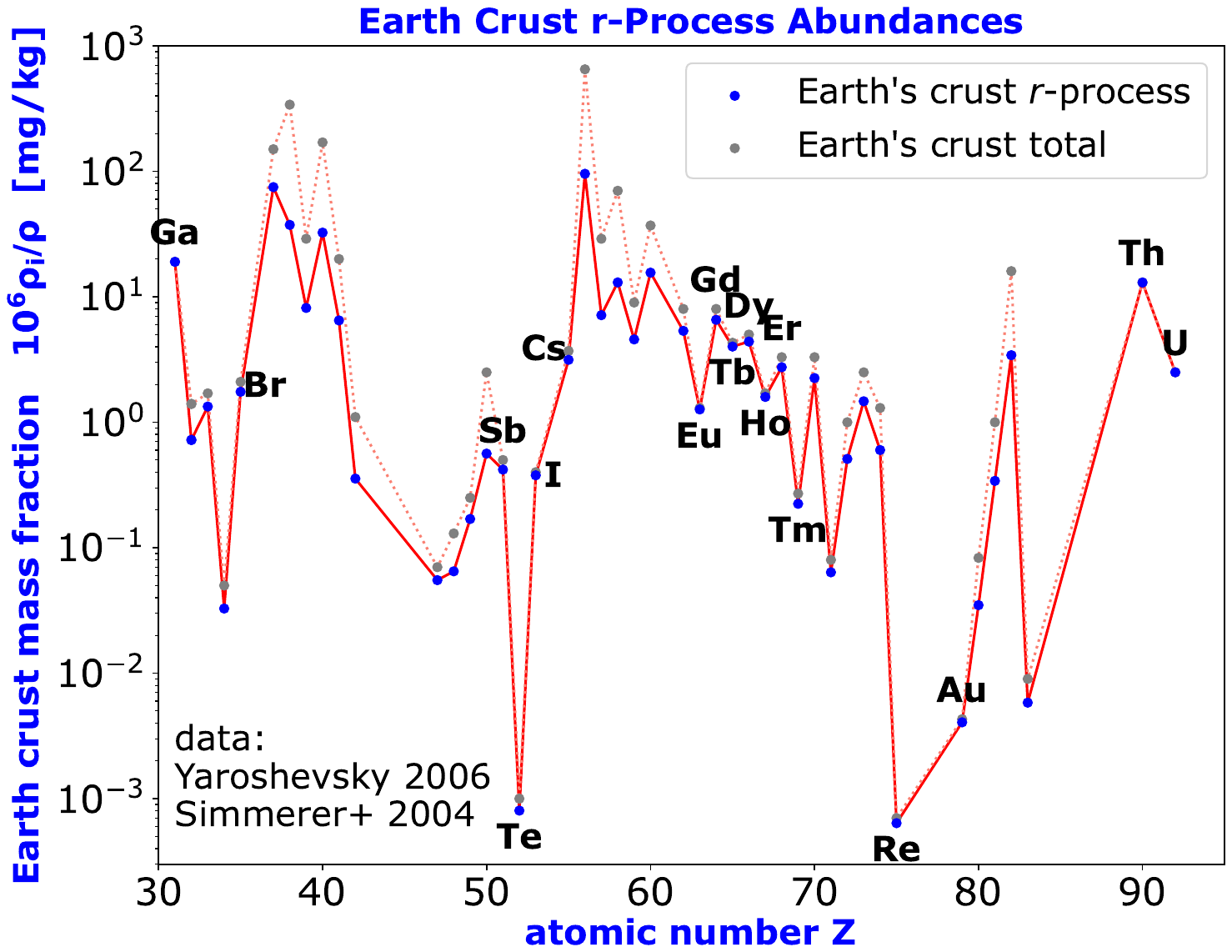}
    \caption{A comparison of measured element abundances in the Earth's crust \cite{yaroshevsky2006abundances}  (dotted line) with
    $r$-process portion (solid line) calculated using the $r$-process fractions from \cite{Simmerer:2004jq}. Elements for which the calculated 
    $r$-process fractions exceed 75\% are labelled. These include the elements
    bromine, iodine, thorium and uranium, whose relevance to life on Earth is
    discussed in the text. 
    }
    \label{fig:rprocess}
\end{figure}

The heavy elements present in the Earth's crust have been produced by a combination of $r$- and $s$-process contributions.
Our calculations of the production of the $^{127}$I that is essential for human life, based on the data of~\cite{Kappler198,Simmerer:2004jq}, indicate that the {\em r}-process has provided 
96\% of its abundance in the Earth's crust, as seen in  Fig.~\ref{fig:rprocess}.  The $r$-process is also calculated to have provided most of the abundance of bromine {and gadolinium} in the Earth's crust, as well as all its thorium and uranium {and a fraction of the molybdenum and cadmium}.

In the following we consider candidate $r$-process sites, with a focus on kilonovae.

\section*{The $r$-Process in Neutron Star Collisions (Kilonovae)}

The $r$-process may occur wherever there is a high {number} of free neutrons {per seed nucleus}, and the nature of the dominant type of site has long been a subject of debate~\cite{thielemann2011astrophysical}. One of the possibilities is in the material ejected during rebound from the core of a core-collapse supernova (CCSN) \cite{Meyer+1992}, as part of supernova nucleosynthesis, and another is in the decompressed neutron-star matter ejected during a neutron star merger (kilonova)~\cite{metzger_kilonovae_2020}.

There is considerable uncertainty in CCSN calculations of the $r$-process, e.g., in the physics of neutrinos and their impact on the neutron abundance \cite{Wang2020} and in the effects of relativistic magnetohydrodynamic jets that can expel neutron-rich material from the proto-neutron star \cite{Reichert+2023}. However, it is particularly difficult in CCSN models to produce the actinides such as thorium and uranium, which may have played in indirect role in the evolution of life on Earth, as discussed above. The presence in deep-ocean deposits of the live isotope $^{244}$Pu (half-life 80 million years) suggests that there has been {$r$-process production of} actinides near Earth within the (cosmologically) recent past~\cite{Wallner2021}. In view of the difficulties with CCSN calculations, we consider the possibility that this {material is attributable to} a neutron star merger (kilonova)~\cite{WangWangXiLu:2021fzc}.

Observations of the kilonova associated with the first binary neutron star merger observed by the LIGO and Virgo experiments to emit gravitational waves (GW170917, see below) indicate that it produced both tellurium and lanthanides as predicted in $r$-process calculations~\cite{Hotokezaka+2023}, and evidence for the production of tellurium and lanthanides by the kilonova associated with GRB 230307A has also been presented~\cite{levan2023heavy}. We recall that tellurium ($Z = 52$) is adjacent to iodine ($Z = 53)$ in the periodic table and in the same abundance peak predicted by the $r$-process, indicating that GW170917 and GRB 230307A probably also produced iodine. This supports the suggestion~\cite{WangWangXiLu:2021fzc} that the kilonova postulated to have produced the terrestrial $^{244}$Pu also produced iodine.

These remarks motivate the following question: Why do neutron stars collide?

\section*{Gravitational Waves Cause Neutron-Star Collisions}

The first, indirect evidence that gravitational waves control the evolution of neutron-star binary systems came from measurements of the binary pulsar PSR B1913+16 that was discovered by Hulse and Taylor in 1974~\cite{Hulse:1974eb}. Following its observation, measurements of the rate of change of the periodicity of electromagnetic pulses from this system over four decades have verified with high accuracy the predictions based on gravitational-wave emission according to the general theory of relativity, which causes energy loss and inspiral of the binary components. The ratio of the observed rate of change compared to the predicted value is $0.9983 \pm 0.0016$~\cite{Weisberg:2016jye}, a verification of the predicted gravitational wave emission at the level of $1.6 \times 10^{-3}$. However, this measurement has now been surpassed by observations of the binary pulsar system PSR J0737-3039 A,B, whose rate of change of periodicity is $0.999963\pm0.000063$ of the value predicted from gravitational wave emission in general relativity~\cite{Kramer:2021jcw}, a verification of the prediction at the level of $6 \times 10^{-5}$.

The first direct evidence for the role of gravitational wave emission in neutron star collisions came with the detection by the LIGO and Virgo experiments of a sequence of gravitational waves of increasing frequency (GW170817) as a pair of neutron stars approached merger~\cite{LIGOScientific:2017vwq}. The rate of increase of the gravitational-wave frequency over an observation period of $\sim 100$~s again agreed perfectly with predictions for gravitational-wave emission during the infall stage of a neutron-star binary system as predicted by general relativity.

We emphasize that these two sets of observations confirm the expected evolution of neutron-star binaries during very different stages of their evolution.  The Hulse-Taylor pulsar is calculated to have a lifetime to merger of $3 \times 10^8$~y, whereas GW170817 was measured during its death throes. The combination of these measurements confirms gravitational-wave domination of the evolution of neutron-star binaries during very different stages of their existence.

Neutron-star collisions are caused by gravitational waves.

\section*{Lunar Tests of Iodine Production by Kilonovae}

The chain of argument presented above suggests that the iodine essential for human life was probably produced by the $r$-process in the collisions of neutron stars that were induced by the emission of gravitational waves, as well as other essential heavy elements. However, there are two weak points in this reasoning: one is that there are other possible sites for the $r$-process such as CCSNe, and the other is that the evidence for iodine production in kilonovae is still only circumstantial.

The iodine essential for human life is the stable isotope $^{127}$I, whose astrophysical origin is challenging to verify directly. 
Astronomical observations of other elements in kilonova explosions, see, e.g.,~\cite{levan2023heavy}, provide a means of verifying that these outburst are an important {\em r}-process  site.
In addition, new astrophysical messengers can now offer complementary information. In particular, the unstable isotope $^{129}$I (half-life 16 million years) from astrophysical sources can in principle be detected using the ultraprecise accelerator mass spectrometry (AMS) technique. This has been used in terrestrial searches, which are unfortunately subject to anthropogenic contamination. Accordingly, we have suggested using AMS to looking for $^{129}$I in conjunction with $^{244}$Pu in samples of lunar regolith (soil)~\cite{Wang:2021dzl}, which have no anthropogenic contamination {\it in situ}.

Model calculations suggest that a kilonova should produce one or two orders of magnitude more $^{129}$I than $^{244}$Pu~\cite{Wang:2021dzl}. However, iodine is volatile, unlike $^{244}$Pu and most other $r$-process radioisotopes of interest, which are refractory. This volatility affects the cosmic dust properties of iodine and hence how it propagates through the interstellar medium. Another factor affecting the relative abundances of $^{129}$I and $^{244}$Pu is the difference in half-lives, which is $\sim 5$: $^{244}$Pu might have been accumulating for period that is longer by a similar ratio. For these reasons, there is considerable uncertainty in the kilonova prediction for the $^{129}$I abundance in the lunar regolith relative to the $^{244}$Pu abundance (which can be estimated from terrestrial measurements~\cite{Wallner2021}). {The latter uncertainty could be mitigated by the simultaneous detection of $^{129}$I and $^{247}$Cm \cite{Cote+2021}.}

A successful search for lunar $^{129}$I would have significant impact: it would provide circumstantial evidence that the $r$-process in kilonovae does indeed produce iodine, suggesting a positive answer to the question in the title of this paper.

\acknow{The work of JE was supported by the United Kingdom STFC Grants ST/X000753/1 and ST/T00679X/1. The work of BDF was supported by National Science Foundation grant AST-2108589.  The work of RS was supported in part by the US Department of Energy Grant DE-FG02-95-ER40934.}

\showacknow{} 

\bibliography{gwlife}

\end{document}